\documentstyle[twoside,fleqn,espcrc2,epsfig]{article}

\newcommand{\beq}{ \begin{equation} }
\newcommand{\eeq}{ \end{equation} }
\newcommand{\bea}{ \begin{eqnarray} }
\newcommand{\eea}{ \end{eqnarray} }

\newcommand{\f}{ \frac }

\newcommand{\simgt}{ \:_\sim^{_>}\, }

\newcommand{\AmS}{{\protect\the\textfont2
  A\kern-.1667em\lower.5ex\hbox{M}\kern-.125emS}}

\title{Fixed point actions in SU(3) gauge theory: surface tension and
  topology \thanks{Presented by A. Papa who is the author of
  Sect.~\ref{sec:therm}. Sect.~\ref{sec:top} was done in collaboration with
  F. Farchioni.}} 

\author{F. Farchioni and A. Papa\address{Institute for Theoretical Physics,
University of Bern \\ Sidlerstrasse 5, CH-3012 Bern, Switzerland}}

\begin{document}
\pagestyle{empty}

\begin{abstract}
This work is organized in two independent parts. In the first part are 
presented some results concerning the surface tension in SU(3) obtained with a
parametrized fixed point action. In the second part,  
a new, approximately scale-invariant, parametrized fixed point action
is proposed which is suitable to study the topology in SU(3). 
\end{abstract}

\maketitle

\section{INTRODUCTION}

Lattice actions living on the direction of the space of couplings which
originates at the fixed point of a renormalization group (RG) transformation
and leaves orthogonally the critical surface, are called fixed point (FP)
actions. They are {\em classical} perfect actions and have many interesting
properties~\cite{Has97}. In particular, FP actions possess scale-invariant
instanton solutions. 

Given a RG transformation, the FP action of any 
lattice configuration is well defined and can be determined numerically by 
multigrid minimization~\cite{HN94}. In Monte Carlo simulations, however, only
simple enough {\em parametrizations} of the FP action can be used.

In Sect.~\ref{sec:therm} the so called ``type III'' parametrized FP action
proposed in Ref.~\cite{BN96} is used to study some typical observables
in SU(3) thermodynamics at high temperatures (free energy density) and at
criticality (surface tension). In Sect.~\ref{sec:top} a new,
approximately scale-invariant parametrization of the FP action is proposed 
to be used in SU(3) topology.

\vspace{-.1cm}
\section{SU(3) THERMODYNAMICS}
\label{sec:therm}

In lattice numerical simulations, the temperature $T$ and 
the volume $V$ are determined by the lattice size $N_\sigma^3\times
N_\tau$, ($N_\tau < N_\sigma$) through $T=1/(N_\tau a)$ and $V=(N_\sigma a)^3$.

At high temperatures, high momentum modes give relevant contributions to
the free energy density. In this regime, thermodynamic 
quantities are strongly influenced by the finite lattice cut-off 
$a^{-1}$, i.e. they show $1/N_\tau^n$ corrections from the
continuum at fixed $T$.
For the Wilson action these cut-off effects are very large:
in the case of the energy density of the ideal gluon gas at $N_\tau=4$
they are as large as 50\%~\cite{BKL95}.
Therefore, the continuum can be extrapolated only from lattice results
at relatively large values of $N_\tau$. 

In Ref.~\cite{Pap96} the free energy density of SU(3) at $T/T_c = 4/3, \ 3/2,
\ 2$ has been determined using the type III parametrized FP action. 
Simulations were performed on lattices as small as  
$8^3\times 2$ and $12^3\times 3$. Results at $N_\tau=3$ showed 
already a good agreement with the continuum extrapolated from the Wilson
action determinations on lattices with $N_\tau=6$ and 8 (see also
Ref.~\cite{Lae97}). 

\begin{table*}[htb]
\setlength{\tabcolsep}{1.1pc}
\catcode`?=\active \def?{\kern\digitwidth}
\caption{Parameters of the runs and results for $\beta_c$ and
  $\chi_L/N_\sigma^3$. One iteration is the combination of a 20-hit
Metropolis and 4 over-relaxation updatings. Errors have been estimated by the
jackknife method.} 
\label{table:s}
\begin{tabular*}{\textwidth}{lccccc}
\hline
  \multicolumn{1}{l}{lattice} 
& \multicolumn{1}{c}{\# $\beta$} 
& \multicolumn{1}{c}{\# iterations} 
& \multicolumn{1}{c}{$\beta$ reweighting} 
& \multicolumn{1}{c}{$\beta_c$} 
& \multicolumn{1}{c}{$\chi_L/N_\sigma^3$}  \\
\hline
$12^3\times3$ &3& 645099 & 3.58995 & 3.58982(9) & $2.958(20)\times10^{-2}$ \\
$12^3\times4$ &1& 171209 & 3.69915 & 3.7007(3) & $1.013(16)\times10^{-2}$ \\ 
$16^3\times4$ &3& 135662 & 3.70025 & 3.7009(5) & $8.87(20)\times10^{-3}$ \\ 
\hline
\end{tabular*}
\vspace{-.6cm}
\end{table*}

At the critical temperature of the first-order deconfinement transition of
SU(3), there can be mixed states where the confined and the deconfined phases 
coexist, separated by an interface. These mixed states have an additional free
energy $F = \sigma A$ ($\sigma$ is the surface tension, $A$ the area of the
interface). The frequency distribution of any order parameter $\Omega$ at the
transition shows a typical double-peak structure, where the two peaks
correspond to the pure phase configurations, while the region in-between
corresponds to configurations containing an interface. The peaks become more
pronounced when the volume is increased. The leading volume dependence of the
surface tension can be determined by~\cite{IKKRY94}
\beq
\left(\f{\sigma}{T_c^3}\right) = - \f{1}{2} \left(\f{N_\tau}{N_{\sigma}}
\right)^2 
\ln\f{P_{min}}{P_{max,1}^{\gamma_1}P_{max,2}^{\gamma_2}}, 
\label{eq:s}
\eeq
with $\Omega = \gamma_1 \Omega_1 + \gamma_2 \Omega_2, \;\;
\gamma_1+\gamma_2=1$. Here $P_{min}$ is the minimum of the distribution
$P(\Omega)$, $P_{max,1}$ and $P_{max,2}$ are the two maxima corresponding to
the values $\Omega_1$ and $\Omega_2$ of $\Omega$ in the pure states of the two
phases at infinite volume.  

\begin{figure}[htb]
\vspace{-1.cm}
\epsfxsize=75mm
\epsfbox{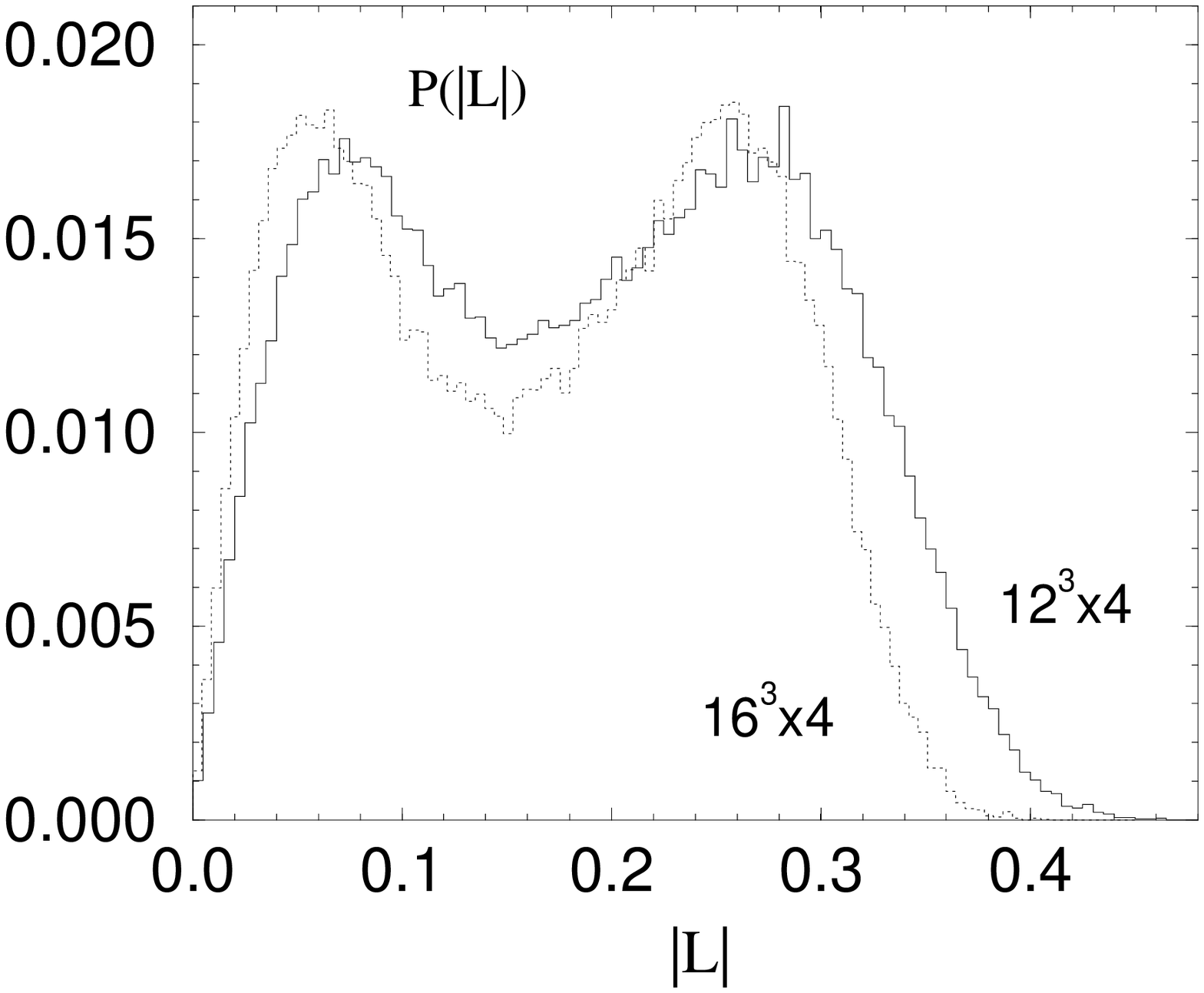}
\vspace{-1.5cm}
\caption{Polyakov loop distribution on lattices $12^3\times4$ and
  $16^3\times4$. Data have been normalized to the total statistics.}  
\label{fig:12x4_16x4}
\end{figure}

A convenient choice for the order parameter in SU(3) is the absolute value of
the Polyakov loop $L = 1/N_\sigma^3 \sum_{\vec{n}}{\rm Tr
  }\prod_{n_4=1}^{N_\tau} U_\mu (\vec{n},n_4)$. Numerical simulations with the
type III parametrized FP action have been performed on three lattices for
values of $\beta$ close to criticality (Table~\ref{table:s} and
Fig.~\ref{fig:12x4_16x4}). Ferrenberg-Swendsen 
reweighting has been applied in order to make the peaks in the Polyakov loop
distribution have the same height. The critical couplings have been determined
through the location of the peak in the Polyakov loop susceptibility
$\chi_L = N_\sigma^3 (\langle |L|^2 \rangle - \langle |L|\rangle^2)$. They are
in agreement within errors with the results of Ref.~\cite{BN96} where the so
called Columbia definition was adopted.
In Table~\ref{table:s_sum} the results for $\sigma/T_c^3$ are compared 
with other determinations using the tree-level Symanzik improved action
(plaquette + rectangle) and the tadpole improved action (with the
same loops)~\cite{BKP96}.\footnote{For comparison with the Wilson action see
Ref.~\cite{IKKRY94}.}  
The infinite volume extrapolations have been done according to the
ansatz~\cite{IKKRY94}  
\beq
\left(\f{\sigma}{T_c^3}\right)_V = \left(\f{\sigma}{T_c^3}\right)
- \left(\f{N_\tau}{N_{\sigma}}\right)^2 \left[c +
\f{1}{4}\ln N_\sigma\right].
\eeq

Table~\ref{table:s_sum} shows that the tadpole improved action has no cut-off 
dependence from $N_\tau=3$ to $N_\tau=4$, thus indicating for $\sigma/T_c^3$ 
a continuum value equal to 0.0155(16) (corresponding to $\sigma\sim 7 \ 
{\rm MeV/fm}^2$). The type III FP action results deviate from this continuum 
value both at $N_\tau=3$ and $N_\tau=4$. This is surprising in consideration
of the results obtained with the same action in the case of the free energy
density.  
The surface tension is, however, a difficult
quantity to determine for the high statistics required and for the
complicated volume dependence. Further checks are necessary before drawing a
definite conclusion.  

\begin{table}[bt]
\vspace{.2cm}
\setlength{\tabcolsep}{0.33pc}
\catcode`?=\active \def?{\kern\digitwidth}
\caption{ $\sigma/T_c^3$ for various lattice actions on several lattices.
Errors estimated by the jackknife method.}
\label{table:s_sum}
\begin{tabular}{cccc}
\hline
  \multicolumn{1}{c}{lattice} 
& \multicolumn{1}{c}{tree-level~\cite{BKP96}} 
& \multicolumn{1}{c}{tadpole~\cite{BKP96}} 
& \multicolumn{1}{c}{type III FP}  \\
\hline
 $12^3\times3$       & 0.0234(24) &  0.0158(11) & 0.0307(8)\\
\hline
 $12^3\times4$       &            &             & 0.0196(11)\\
 $16^3\times4$       & 0.0148(16) &  0.0147(14) & 0.0180(21)\\
 $24^3\times4$       & 0.0136(25) &  0.0119(21) & \\
 $32^3\times4$       & 0.0116(23) &  0.0125(17) & \\
\hline
 $\infty^3\times4$   & 0.0152(26) &  0.0152(20) & 0.026(6)\\
\hline
\end{tabular}
\vspace{-.8cm}
\end{table}

\vspace{-.8cm}
\section{TOPOLOGY}
\label{sec:top}

It is well known that FP actions possess scale-invariant instanton
solutions~\cite{HN94}: the dependence of the FP action on the instanton size
$\hat\rho$ is flat up to $\hat\rho$ as small as $\sim 1$ lattice
spacing~\cite{BBHN96}. 
FP actions combined with a suitable definition of the topological
charge~\cite{BBHN96,DFP97,Nie97} allow a consistent determination of the
topological susceptibility. 

\begin{table}[bt]
\setlength{\tabcolsep}{0.33pc}
\catcode`?=\active \def?{\kern\digitwidth}
\caption{Couplings of the parametrized FP action.}  
\label{table:par}
\begin{tabular}{ccccc}
\hline
  \multicolumn{1}{c}{loop} 
& \multicolumn{1}{c}{$c_1$} 
& \multicolumn{1}{c}{$c_2$} 
& \multicolumn{1}{c}{$c_3$} 
& \multicolumn{1}{c}{$c_4$}  \\
\hline
plaquette  & --0.5441 &   1.9405 & --0.4663 &   0.0162 \\
bent rect. &   0.1099 & --0.1150 & --0.0087 &   0.0073 \\
twisted-8  & --0.0089 &   0.0048 &   0.0154 & --0.0014 \\
\hline
\end{tabular}
\vspace{-.6cm}
\end{table}

In this Section we present a new, approximately scale-invariant {\em
parametrization} of the type III FP action for SU(3), determined by a fit
procedure on the (numerically estimated) ``exact'' values of the FP action of
both typical Monte Carlo configurations and ``hand-made'' instanton solutions
with $\hat\rho\simgt1$. The procedure to build such instantons was  
first described in Ref.~\cite{BBHN96} for the O(3) $\sigma$ model and
applied to SU(2) in Ref.~\cite{DHZ96}. Differently from Ref.~\cite{DHZ96}, we
bypassed the problem of non-existence of single instantons on lattices with
periodic b.c. by working with {\em open} boundaries. \newline 
We performed the following steps: \newline
-- we discretized smooth continuum instantons on a $62^4$ lattice, in order to
have configurations on which Wilson action and typical Symanzik actions are the
same within a few per mill; \newline
-- we blocked down three times by the type III RG averaging, in order to
make the final instanton size $\hat\rho\sim1$; \newline
-- we estimated by numerical minimization the ``exact'' FP action 
of the blocked configurations.

The blocked configurations were included finally in the parametrization
procedure together with $\sim500$ typical Monte Carlo configurations in the
range $\beta_{\rm Wilson}=5.1-50. $\footnote{These Monte Carlo configurations
were obtained in Ref.~\cite{BN96}.}. We searched for parametrizations
of the form  
\beq
A(U) = \f{1}{N}\sum_{C, \ i\geq1} c_i(C)[N-{\rm ReTr}(U_C)]^i ,
\label{eq:par}
\eeq 
where $C$ denotes any closed path, $U_C$ stands for $\Pi_C U_\mu(n)$. 
Skipping the many technical subtleties induced by the open b.c. in the full
procedure and the details of the fit, we just mention that we defined the
action on a finite lattice $\Lambda$, 
$A_\Lambda = \sum_{x\in\Lambda} A(x)$, through 
\beq
A(x) = \f{1}{N} \sum_{C\ni x, i\geq1} c_i(C) \f{[N-{\rm ReTr}(U_C)]^i}{{\rm
  perimeter}(C)} \;\;\;.  
\eeq

\begin{figure}[htb]
\vspace{-.8cm}
\epsfxsize=75mm
\epsfbox{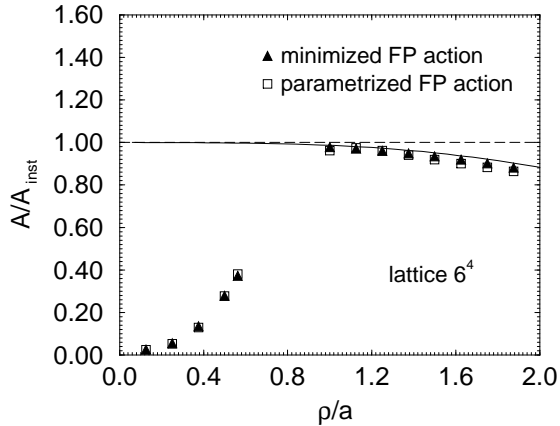}
\vspace{-1.5cm}
\caption{Minimized and parametrized FP action of the blocked
  instanton configurations versus the instanton size $\hat\rho$. The solid
  line is the finite volume instanton action on the continuum.}
\label{fig:par}
\end{figure}

We found a nice parametrization involving the plaquette, the bent rectangle
and the twisted perimeter-8 loop (x,y,z,t,-x,-y,-z,-t), with the couplings
given in Table~\ref{table:par}. In Fig.~\ref{fig:par} parametrized and
minimized FP action are compared with the finite volume instanton action on
the continuum.

\vspace{-.8cm}
\begin{thebibliography}{9}
\bibitem{Has97} P. Hasenfratz, these proceedings.
\bibitem{HN94} P. Hasenfratz and F. Niedermayer, Nucl. Phys. B414 (1994)
  785.
\bibitem{BN96} M. Blatter and F. Niedermayer, Nucl. Phys. B482 (1996)
  286. 
\bibitem{BKL95}  B. Beinlich, F. Karsch and E. Laermann, Nucl. Phys. B462
  (1996) 415. 
\bibitem{Pap96} A. Papa, Nucl. Phys. B478 (1996) 335.
\bibitem{Lae97} E. Laermann, these proceedings.
\bibitem{IKKRY94} Y. Iwasaki et al., Phys. Rev. D49 (1994) 3540.
\bibitem{BKP96} B. Beinlich, F. Karsch and A. Peikert, Phys. Lett. B390 
(1997) 268.
\bibitem{BBHN96} M. Blatter, R. Burkhalter, P. Hasenfratz and F. Niedermayer, 
Phys. Rev. D53 (1996) 923.
\bibitem{DFP97} M. D'Elia, F. Farchioni and A. Papa, Phys. Rev. D55 (1997)
  2274. 
\bibitem{Nie97} F. Niedermayer, Nucl. Phys. B (Proc. Suppl.) 53 (1997) 56.
\bibitem{DHZ96} T. DeGrand, A. Hasenfratz and D. Zhu, Nucl. Phys. B475 (1996)
  321. 
\end{thebibliography}
\end{document}